\documentclass[%
aip,
cp,  
 amsmath,amssymb,
 reprint,%
]{revtex4-2}

\usepackage{graphicx}
\usepackage{dcolumn}
\usepackage{bm}

\usepackage[utf8]{inputenc}
\usepackage[T1]{fontenc}
\usepackage{mathptmx} 
\usepackage{amsmath} 
\newcommand{\mean}[1]{\left\langle #1 \right\rangle}
\begin{document}

\title{Comparison of Methods for Elliptic Flow Measurements at NICA Energies $\sqrt{s_{NN}}$ = 4 -- 11 GeV}

\author{Vinh Ba Luong} 
 \email[Corresponding author: ]{lbavinh@gmail.com}
\author{Dim Idrisov} %
\author{Petr Parfenov}
\author{Arkadiy Taranenko}
\author{Alexander Demanov}
\affiliation{
  National Research Nuclear University MEPhI (Moscow Engineering Physics Institute), \\ Kashirskoe highway 31, Moscow, 115409, Russia.
}

\date{\today} 
%
\begin{abstract}
  The goal of the Multi-Purpose Detector (MPD) experiment at NICA collider is to
  explore the QCD phase diagram of strongly interacting matter
  produced in nucleus-nucleus collisions at $\sqrt{s_{NN}}=4-11$~GeV. 
  The performance of MPD detector  for elliptic
  flow measurements of  charged hadrons
  is studied with Monte-Carlo
  simulations using collisions of Au+Au
  ions employing UrQMD, SMASH, and AMPT heavy-ion event generators. Different methods for
  flow measurements: event plane and direct cumulants are used to investigate
  the contribution of non-flow correlations and flow fluctuations.
\end{abstract}
\maketitle
\section{Introduction}
The Multi-Purpose Detector (MPD) experiment will be one of the main scientific
pillars of the future  Nuclotron-based Ion Collider fAcility (NICA) at JINR, Dubna \cite{nica,mpd}. The
main goal of the MPD research program is to explore the QCD phase diagram in the region of
high baryon densities using relativistic heavy-ion collisions at $\sqrt{s_{NN}}=4-11$~GeV.
The anisotropic collective flow is one of the most important observables
sensitive to the transport properties of the strongly interacting matter and it can be quantified by the 
Fourier coefficients $v_n$ in the expansion of the particles azimuthal distribution 
as: $dN/d\phi \propto 1 + \sum_{n=1} 2 v_{n} \cos 
(n(\phi-\Psi_{n}))$~\cite{flow2}, where $n$ is the order of 
the harmonic, $\phi$ is the azimuthal angle of a particle of a given type, 
and $\Psi_n$ is the azimuthal angle of the $n$th-order event plane. Elliptic flow, $v_{2} = \langle{\cos[2(\phi - \Psi_n)]}\rangle$,
is the dominant flow signal at NICA energy regime. In this work, we discuss the anticipated physics performance of 
MPD detector system for elliptic flow measurements  at NICA energies.
\section{The MPD detector system at NICA}
\begin{figure}[htb]
	\centering
	\includegraphics[width=0.45\textwidth] {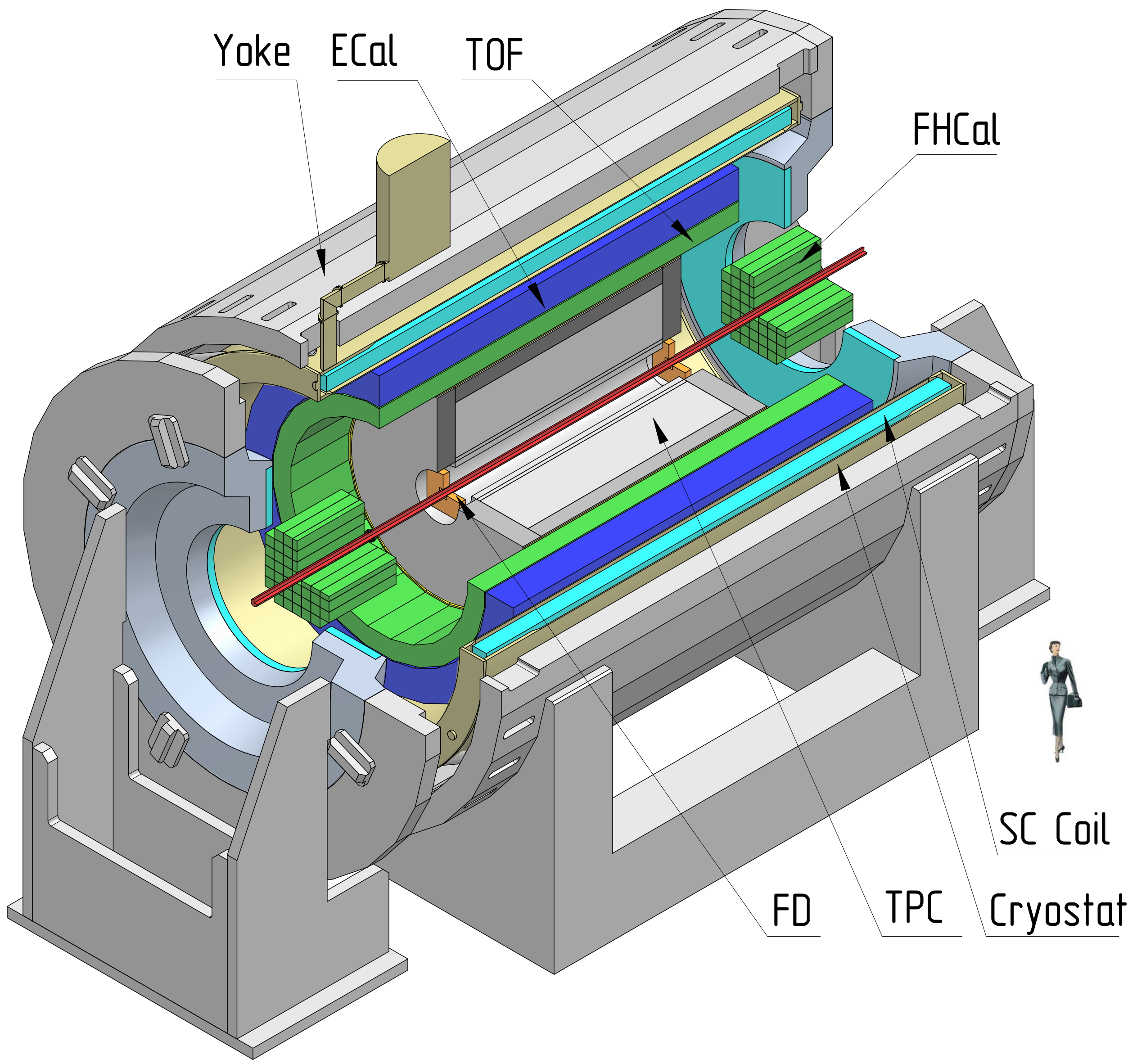}
	\includegraphics[width=0.45\textwidth] {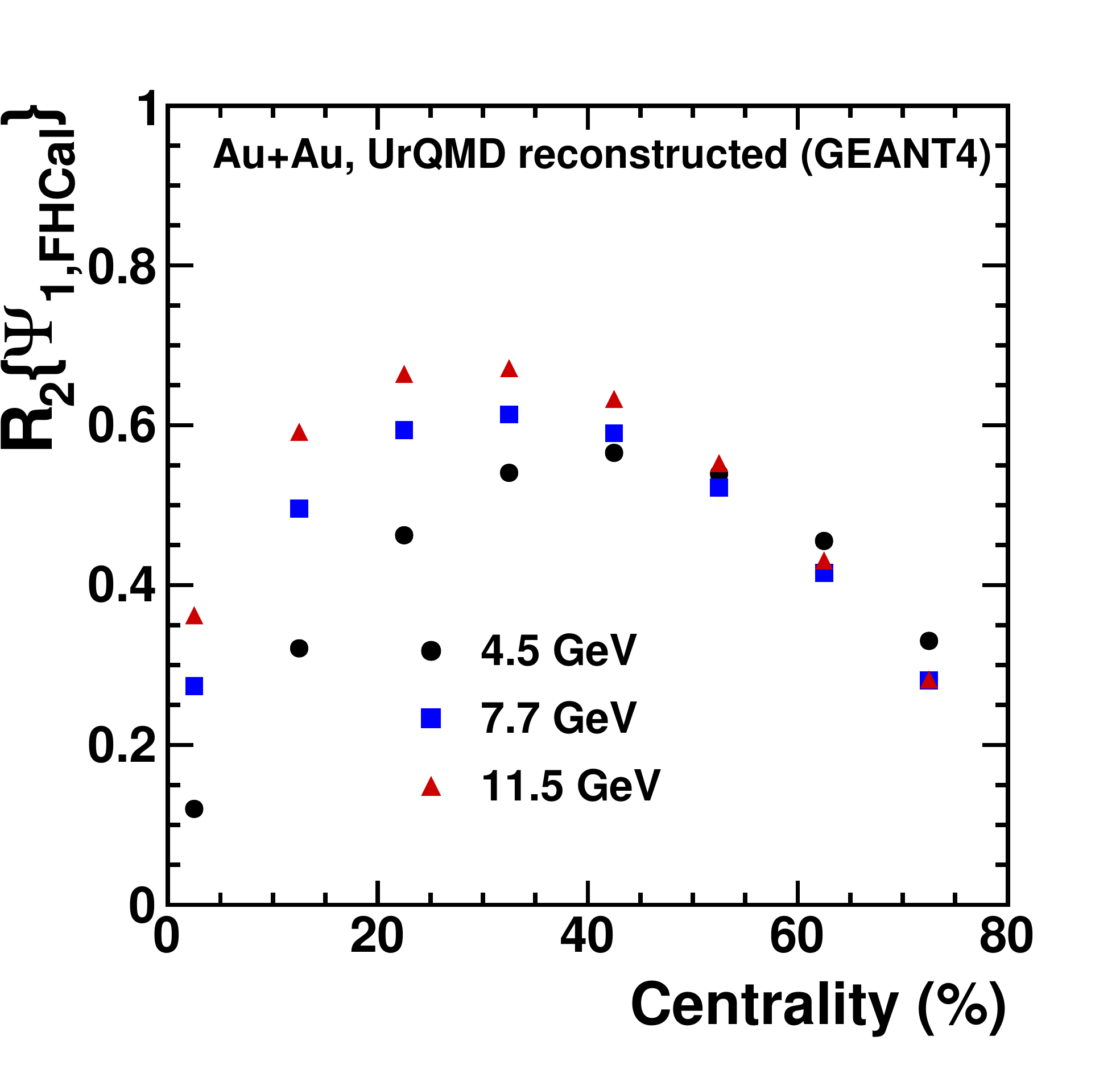}
	\caption{(left) The schematic view of the MPD detector in Stage 1. (right)
          Centrality dependence of event plane resolution factor $\rm R_2(\Psi_{1,\textrm{FHCal}})$ (right)
          for Au+Au collisions at $\sqrt{s_{NN}}$ = 4.5, 7.7, and 11.5 GeV.}
	\label{fig:res}
\end{figure}

The MPD detector system (Fig.~\ref{fig:res}, left) consists of a barrel part and two endcaps located inside the magnetic field. Time Projection Chamber (TPC) will be the central MPD tracking detector \cite{mpd}. TPC will provide 3D tracking of charged particles, as well as the
measurement of specific ionization energy loss $dE/dx$ for particle identification for $|\eta|<$ 1.5.
The TPC will be surrounded by a cylindrical barrel of the Time-of-Flight (TOF) detector with a timing resolution of the order of 50 ps. The combined system TPC+TOF will allow the efficient charged pion/kaon separation up to 1.5 GeV/c and protons/meson separation up to 2.5 GeV/c.
The Forward Hadronic Calorimeter
(FHCal), placed at 2 $<|\eta|<$ 5, will be used for centrality determination as well
for the reconstruction of event plane from the directed flow of particles.

In this work, we use the cascade version of  UrQMD~\cite{Bleicher:1999xi,Bass:1998ca},
SMASH~\cite{Weil:2016zrk}, and string melting version of AMPT~\cite{Lin:2004en} models
to simulate the heavy-ion collisions at NICA energies.
In total, the sample of 120 M of minimum bias Au+Au events at $\sqrt{s_{NN}}=7.7$ and 11.5 GeV was
used for elliptic flow performance study using different methods of analysis. We used the
term ``true'' $v_2$ data for these results. At the next step, a sample of
20 M UrQMD minimum bias events was used as an input for the full chain of the realistic simulations of the MPD detector subsystems based on the GEANT4 platform and reconstruction algorithms built in the MPDROOT.
We named these $v_2$ results the ``reco'' $v_2$ data.
The main workflow for the analysis of identified charged hadrons with the
reconstructed data is similar to  the previous work~\cite{Parfenov:2019pxf}.
\vspace{-0.5pc}
\section{Methods for elliptic flow measurements in MPD}
In this section, we discuss how the event plane, scalar product and direct cumulant methods can be used for the measurements of elliptic flow of the produced particles with MPD detector system at NICA.

The event plane method correlates the azimuthal angle $\phi$ of each particle
with the azimuthal angle $\Psi_n$ of event plane determined from the anisotropic
flow itself \cite{vol2008}.  The event flow vector ($Q_n$) and the azimuthal angle of event plane $\Psi_n$ can be defined for each harmonic, $n$, of the Fourier expansion by:
\begin{eqnarray}
	Q_{n,x} = \sum \limits_{i} \omega_{i} \cos (n\varphi_i),\ Q_{n,y} = \sum \limits_{i} \omega_{i} \sin (n\varphi_i), \Psi_n = \frac{1}{n} \tan^{-1} \left( \frac{Q_{n,y}}{Q_{n,x}} \right),
\end{eqnarray}
where the sum runs over all particles $i$ used in the event plane calculation, and $\varphi_i$  and  $\omega_i$ are the laboratory azimuthal angle and the weight for the particle $i$. 

For $v_2$ measurements in MPD  we can use both event planes  from
elliptic ($\Psi_{2,\textrm{TPC}}$) and directed flow ($\Psi_{1,\textrm{FHCal}}$):
\begin{equation}
v_{2}\{\Psi_{2,\text{TPC}}\}=\frac{\langle\cos(2(\phi_{i} - \Psi_{2,\text{TPC}}))\rangle}{R_{2}(\Psi_{2,\text{TPC}})}, \:
v_{2}\{\Psi_{1,\text{FHCal}}\}=\frac{\langle\cos(2(\phi_{i} - \Psi_{1,\text{FHCal}}))\rangle}{R_{2}(\Psi_{1,\text{FHCal}})}, 
\end{equation}
where $\rm R_2(\Psi_{2,\textrm{TPC}})$ and $\rm R_2(\Psi_{1,\textrm{FHCal}})$ represent the resolution of the event planes. Here, the $\Psi_{1,\textrm{FHCal}}$ determined from the
directed flow ($n=$ 1) of particles detected in the FHCal (2 $<|\eta|<$ 5) and $\Psi_{2,\textrm{TPC}}$
determined from the elliptic flow ($n=$ 2) of produced particles detected in the TPC ($|\eta|<$ 1.5).
As an example, the right panel of Fig~\ref{fig:res} shows the centrality dependence of 
$\rm R_2(\Psi_{1,\textrm{FHCal}})$ for Au+Au collisions at $\sqrt{s_{NN}}$ = 4.5, 7.7, and 11.5 GeV. The results are
based on the analysis of the fully reconstructed UrQMD events.

In the scalar product method (SP), for differential flow $v_2(p_T)$ measurements, one uses 
the magnitude of the flow vector ($Q_2$) as a weight \cite{vol2008}:
\begin{eqnarray}
	\label{eq:flow_SP}
	v_2^\textrm{SP}\{Q_{2,\textrm{TPC}}\}(p_T)
        = \left\langle u_{2,i}(p_T) Q_2^* \right\rangle /2\sqrt{\left\langle Q_2^a Q_2^{b*} \right\rangle },
\end{eqnarray}
where $u_{2,i}$ is the unit vector of the $i^{th}$ particle (which is not included in $Q_2$ vector) and $a$
and $b$ are two subevents. If $Q_2$ vector is replaced by its unit vector, the scalar product method reduces to
event plane method.

In the Q-cumulant method the two- and four-particle cumulants (for each harmonic $n$)
can be calculated directly from a $Q_n$ vector, constructed
using particles from the TPC acceptance $|\eta|<$ 1.5,
$Q_n \equiv \sum_{i}^{M} exp \left( i n \varphi_i \right)$~\cite{Bilandzic:2010jr}:
\begin{gather}
	\left\langle 2 \right\rangle_n 
	= (\left| Q_n \right|^2 - M) /M(M-1),
	\\
	\left\langle 4 \right\rangle_n 
	= \frac{\left| Q_n \right|^4 + \left| Q_{2n} \right|^2 - 2 \Re [Q_{2n}Q_n^*Q_n^*] -4(M-2)\left| Q_n \right|^2 - 2M(M-3) }{M(M-1)(M-2)(M-3)},
\end{gather}
where $M$ denotes the multiplicity in each
event used in the analysis. The elliptic flow ($n=2$) can be defined via the Q-cumulant method as follows:
\begin{eqnarray}
	v_2\{2\} = \sqrt{\left\langle \left\langle 2 \right\rangle \right\rangle},\ v_2\{4\} = \sqrt[4]{2\left\langle \left\langle 2 \right\rangle \right\rangle^2 - \left\langle \left\langle 4 \right\rangle \right\rangle},
\end{eqnarray}
where the double brackets denote weighted average over all events. Equations for the $p_T$-differential elliptic flow can be
found in~\cite{Bilandzic:2010jr}.

Different methods of analysis can be affected by nonflow and flow fluctuations in different ways. The nonflow effects are mainly due to few particle correlations, not associated with the
reaction plane: Bose-Einstein correlations, resonance decays,
momentum conservation. The estimates of $v_2$ 
based on multi-particle cumulants have the advantage of significant
reduction of contribution $\delta_2$ from
nonflow effects: $\left\langle 2 \right\rangle_2 = v_2^2 + \delta_2,\ \left\langle 4 \right\rangle_2 = v_2^4 + 4 v_2^2\delta_2 + 2\delta_2^2$. In order to suppress nonflow effects in $v_2$ results from two-particle correlation methods, one can use rapidity gaps between correlated particles. For
$v_2\{\Psi_{2,\textrm{TPC}}\}$, $v_2^\textrm{SP}\{Q_{2,\textrm{TPC}}\}$, $v_2\{2\}$ we use the
$\eta$-gap of $\Delta\eta >$ 0.1 between the two sub-events. The  $v_2\{\Psi_{1,\textrm{FHCal}}\}$ results
are expected to be less affected by
nonflow due to larger  $\eta$-gap between particles in TPC and FHCal:
$\Delta\eta >$ 0.5.
\section{Elliptic flow fluctuations}
Anisotropic flow can fluctuate event to event. We define the elliptic flow fluctuations by
$\sigma_{v2}^2 = \mean{v_2^2}-\mean{v_2}^2$. Here, the resulting flow signal,
averaged over all events is denoted as $\left\langle v_2 \right\rangle$.
In the case of the Q-cumulants ($v_2\{2\}$ and $v_2\{4\}$), for a Gaussian model of fluctuations  and in the limit 
$\sigma_{v2}\ll \mean{v_2}$, one can write~\cite{Voloshin:2007pc,vol2008}:
\begin{eqnarray}
	\label{eq:fluctuation_cumulants}
	v_2\{2\} = \left\langle v_2 \right\rangle + 0.5\cdot\sigma_{v2}^2/\left\langle v_2 \right\rangle,\ v_2\{4\} = \left\langle v_2 \right\rangle - 0.5\cdot\sigma_{v2}^2/\left\langle v_2 \right\rangle.
\end{eqnarray}
Consequently, one can exploit this difference to investigate flow fluctuations by the ratio $v_2\{4\}/v_2\{2\}$.
A large contribution from flow fluctuations will result in $v_2\{4\}/v_2\{2\}\ll 1$,
while weak one gives $v_2\{4\}/v_2\{2\}\sim 1$.

One of the important sources of $v_2$ flow fluctuations is the participant eccentricity fluctuations in the initial geometry of the overlapping region of two colliding nuclei. Therefore, the $v_2\{\Psi_{1,\textrm{FHCal}}\}$ values are expected to  be smaller than $v_2\{\Psi_{2,\textrm{TPC}}\}$ measured with respect to the participant plane $\Psi_{2,\textrm{TPC}}$~\cite{Voloshin:2007pc,vol2008}:
\begin{eqnarray}
	\label{eq:fluctuation_EP}
	v_2\{\Psi_{1,\textrm{FHCal}}\} \simeq \left\langle v_2 \right\rangle,\ v_2\{\Psi_{2,\textrm{TPC}}\} \simeq \left\langle v_2 \right\rangle + 0.5\cdot\sigma_{v2}^2/\left\langle v_2 \right\rangle.
\end{eqnarray}

It should be noted that there are similar studies on elliptic flow in NICA energy regime~\cite{Petersen:2006vm, Petersen:2009vx} and analysis of non-flow effects and fluctuations~\cite{Zhu:2005qa}.
\begin{figure}[htb]
	\centering
	\includegraphics[width=0.96\textwidth] {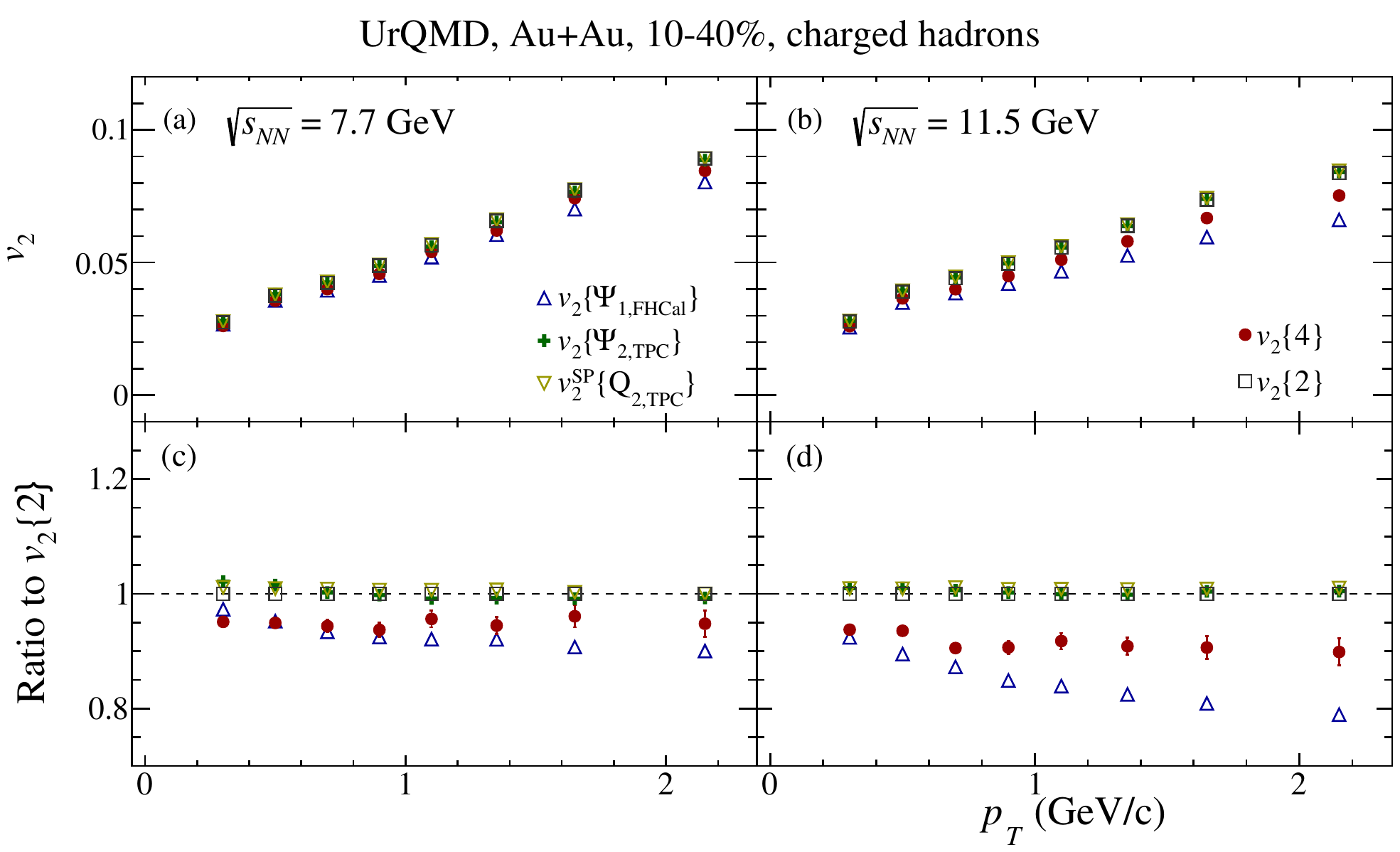}
	\vspace{-0.5pc}
	\caption{ $p_T$-dependence of $v_2$ of inclusive charged hadrons  from
          10-40\% mid-central Au+Au collisions at $\sqrt{s_{NN}}=7.7$~GeV and 11.5 GeV obtained using the  event
          plane ($v_2\{\Psi_{1,\textrm{FHCal}}\}$, $v_2\{\Psi_{2,\textrm{TPC}}\}$), scalar product
          $v_2^{SP}\{Q_{2,\textrm{TPC}}\}$) and Q-cumulant
          ($v_2\{2\}$, $v_2\{4\}$) methods.
          Lower row shows the ratio $v_2$(method)/$v_2\{2\}$.}
	\label{fig:methods_compare}
\end{figure}
\section{Results}
Figure~\ref{fig:methods_compare} shows the $p_T$ dependence of $v_2$ of inclusive charged hadrons
from 10-40\% mid-central Au+Au collisions at $\sqrt{s_{NN}}=7.7$~GeV and 11.5~GeV.
Different symbols correspond to the $v_2$ results obtained by event plane ($v_2\{\Psi_{1,\textrm{FHCal}}\}$, $v_2\{\Psi_{2,\textrm{TPC}}\}$), scalar product
$v_2^{SP}\{Q_{2,\textrm{TPC}}\}$) and Q-cumulant ($v_2\{2\}$, $v_2\{4\}$) methods of
analysis of events from UrQMD model. The ratios of $v_2$ signal to the $v_2\{2\}$ are
shown on the bottom panels and show good agreement between $v_2$ results obtained
by $v_2\{\Psi_{2,\textrm{TPC}}\}$, $v_2^{SP}\{Q_{2,\textrm{TPC}}\}$ and  $v_2\{2\}$ methods.
Both $v_2\{4\}$ and $v_2\{\Psi_{1,\textrm{FHCal}}\}$ methods give a smaller $v_2$ signal as one expects from elliptic flow fluctuations and nonflow effects. The difference is larger for $\sqrt{s_{NN}}=11.5$~GeV than
for $\sqrt{s_{NN}}=7.7$~GeV.

\begin{figure}[htb]
	\centering
	\includegraphics[width=0.9\textwidth] 
	{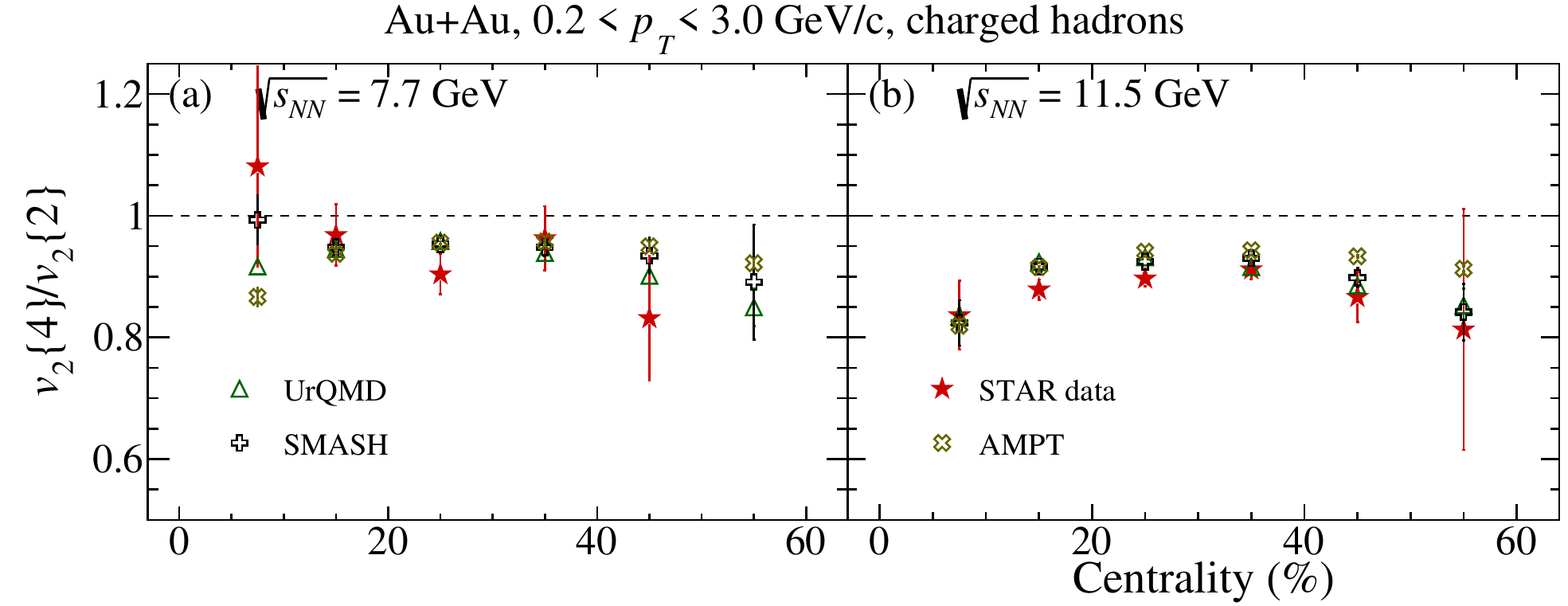}
	\caption{Centrality dependence of relative elliptic flow fluctuations $v_2\{4\}/v_2\{2\}$ of charged hadrons in Au+Au collisions at $\sqrt{s_{NN}}=7.7$ (a) and $11.5$ (b) GeV using UrQMD, SMASH and AMPT models compared to the STAR published data~\cite{Adamczyk:2012ku}.}
	\label{fig:Flow_ratio_comparison}
\end{figure}

The comparison of the ratio $v_2\{4\}/v_2\{2\}$ is shown in Fig.~\ref{fig:Flow_ratio_comparison} between the STAR published data~\cite{Adamczyk:2012ku}, UrQMD, SMASH, and AMPT models for Au+Au collisions at $\sqrt{s_{NN}}=7.7$ (a) and $11.5$~(b) GeV. The results show both hadronic cascade (UrQMD, SMASH) and multi-phase (AMPT) models reproduce the ratio $v_2\{4\}/v_2\{2\}$ from the experimental data reasonably well. In addition, the magnitude of the elliptic flow fluctuations decreases (corresponding with the increase of the ratio $v_2\{4\}/v_2\{2\}$) from central to mid-central collisions. These results are consistent with the expectation that the dominant source of flow fluctuations is the initial eccentricity fluctuations.

\begin{figure}[htb]
	\centering
	\includegraphics[width=0.96\textwidth] {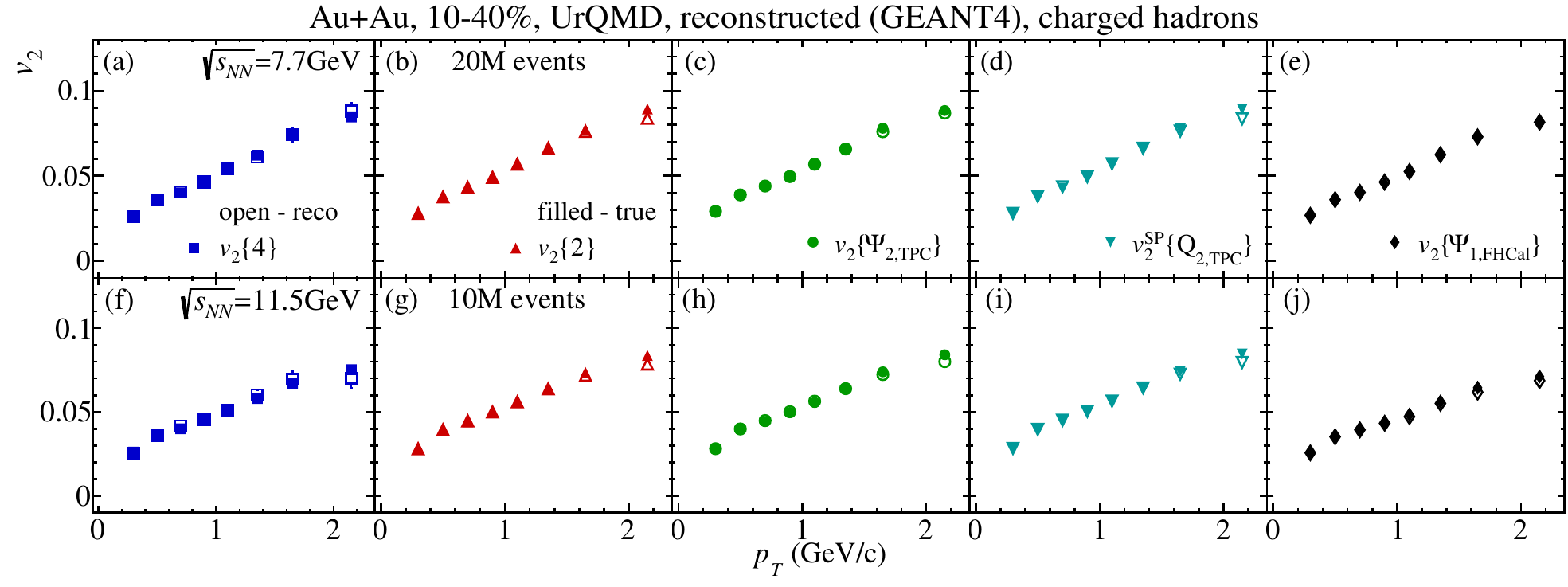}
	\vspace{-0.5pc}
	\caption{ Comparison of $v_2(p_T)$ for charged hadrons from
          10-40\% mid-central  Au+Au collisions at $\sqrt{s_{NN}}=7.7$~GeV (upper panels) and
          $\sqrt{s_{NN}}=11.5$~GeV (lower panels)
          obtained  by Q-cumulant, event plane, scalar product  methods of analysis of 
         fully reconstructed  ("reco") and generated  UrQMD events ("true").}
	\label{fig:performance}
\end{figure}

The event plane ($v_n\{\Psi_{1,\textrm{FHCal}}\}$, $v_n\{\Psi_{2,\textrm{TPC}}\}$) and Q-cumulant
($v_n\{2\}$, $v_n\{4\}$) methods 
were implemented in the MPDROOT framework.  Figure~\ref{fig:performance} shows the $p_T$ dependence
of $v_2$ of charged hadrons  from 10-40\% mid-central  Au+Au collisions at $\sqrt{s_{NN}}=7.7$~GeV
(upper panels) and 
$\sqrt{s_{NN}}=11.5$~GeV (lower panels).  The perfect agreement between
$v_2$ results from the analysis of fully reconstructed ("reco") and generated ("true") UrQMD events is observed.
\section{Summary}
The MPD detector system's performance
for the elliptic flow measurements $v_2$ of charged hadrons
is studied with Monte-Carlo simulations using collisions of Au+Au ions employing UrQMD, SMASH, and AMPT
heavy-ion event generators.
We have shown how the various experimental measures
of elliptic flow are affected by fluctuations and nonflow at NICA energies.
The comparison of relative elliptic flow fluctuations represented
by the ratio $v_2\{4\}/v_2\{2\}$ of UrQMD, SMASH, AMPT models with the
STAR published data at $\sqrt{s_{NN}}=7.7$ and $11.5$~GeV shows the good agreement. This may indicate that the dominant source of flow fluctuations is the initial eccentricity fluctuations.
The detailed comparison of the $v_2$
results obtained from the analysis of the fully reconstructed data and
generator-level data allows to conclude that MPD system will allow reconstruction of
$v_2$ coefficients with high precision. 

\section{Acknowledgments}
This work is supported by the RFBR according to the research project No. 18-02-40086, the
European Union's Horizon 2020 research and innovation program under grant agreement No. 871072, by the Ministry of Science and Higher Education of the Russian Federation, Project ``Fundamental properties of elementary particles and cosmology'' No 0723-2020-0041.

\end{document}